\documentclass[
prb,twocolumn,
superscriptaddress,showpacs,floatfix,notitlepage]{revtex4-2}
\usepackage{graphicx}% Include figure files
\usepackage{amsmath,amssymb}
\usepackage{color}
\usepackage{url}
\usepackage{hyperref}% add hypertext capabilities
\usepackage{subfigure}
\usepackage{dcolumn}% Align table columns on decimal point
\usepackage{bm}% bold math
\usepackage{placeins}
\usepackage[T1]{fontenc}
\newif\iffigures
\figurestrue

\def\Re{{\rm Re}}
\def\Im{{\rm Im}}

\def\rr{\mathbf{r}}
\newcommand{\angstrom}{\textup{\AA}}
\DeclareMathOperator\arctanh{arctanh}

\begin{document}

\title{Excitonic effects in time-dependent density functional theory from zeros of the density response}

\author{D. R. Gulevich} 
\affiliation{School of Physics and Engineering, ITMO University, St. Petersburg 197101, Russia}

\author{Ya. V. Zhumagulov} 
\affiliation{University of Regensburg, Regensburg, 93040, Germany}

\author{V. Kozin} 
\affiliation{Department of Physics, University of Basel, Klingelbergstrasse 82, CH-4056 Basel, Switzerland}

\author{I. V. Tokatly}
\affiliation{Nano-Bio
  Spectroscopy group and European Theoretical Spectroscopy Facility (ETSF), Departamento de Pol\'imeros y Materiales Avanzados: F\'isica, Qu\'imica y Tecnolog\'ia, Universidad del
  Pa\'is Vasco, Av. Tolosa 72, E-20018 San Sebasti\'an, Spain}
 \affiliation{IKERBASQUE, Basque Foundation for Science, 48009 Bilbao, Spain}
\affiliation{Donostia International Physics Center (DIPC), 20018 Donostia-San Sebasti\'{a}n, Spain}
\affiliation{School of Physics and Engineering, ITMO University, St. Petersburg 197101, Russia}

\date{\today} % date may be explicitly specified

\begin{abstract} 
We show that the analytic structure of the dynamical xc kernels of semiconductors and insulators can be sensed in terms of its poles which mark physically relevant frequencies of the system where the counter-phase motion of discrete collective excitations occurs: if excited, the collective modes counterbalance each other, making the system to exhibit none at all or extremely weak density response. This property can be employed to construct simple and practically relevant approximations of the dynamical xc kernel for time-dependent density functional theory (TDDFT). Such kernels have simple analytic structure, are able to reproduce dominant excitonic features of the absorption spectra of monolayer semiconductors and bulk solids, and promise high potential for future uses in  efficient real-time calculations with TDDFT.
\end{abstract}

\maketitle
{\let\newpage\relax\maketitle}

%\tableofcontents

%--------------------------------------------------------------------------------------------
\section{Introduction}
%--------------------------------------------------------------------------------------------

%The dynamical exchange-correlation (xc) kernels in the time-dependent density functional theory (TDDFT) are commonly treated as a black box introduced to account for the differences in the time-dependent response of the exact interacting system and its approximation within the Hartree mean field theory. Nevertheless, here 

The time-dependent density functional theory (TDDFT) aiming at extending the density functional theory to the description of electronic excitations and electron dynamics, while being in principle exact theory, provides a practically useful alternative to many-body perturbation methods~\cite{TDDFTbyUllrich}. While the TDDFT owes its popularity in the computational condensed-matter physics and computational chemistry to the adiabatic local density approximation (ALDA), the description of excitonic effects has become a serious challenge which provoked a genuine interest of theorists~\cite{Reining-2002,Kim-2002,Marini-2003,Sottile-PRL-2003,Botti-2004,Botti-2007,Turkowski-2009,Sharma-2011,Nazarov-2011,Yang-PRB-2013,Ullrich-Yang-2014,Rigamonti-2015,Berger-PRL-2015,Byun-2017,Cavo-2020,Berger-2020,Suzuki-2020}. Yet early~\cite{Reining-2002,Kim-2002} it has been understood that the account of the long-range Coulomb-like tail~\cite{GGG-1997} of the exchange-correlation (xc) kernels which is missing in the local xc kernels such as ALDA and GGA is crucial for capturing the excitonic effects. Many subsequent works focused on designing a suitable approximation for xc kernel with suitable long-range behaviour~\cite{Marini-2003,Sottile-PRL-2003,Stubner-2004,Botti-2004,Botti-2007,Turkowski-2009,Sharma-2011,Nazarov-2011,Yang-PRB-2013,Ullrich-Yang-2014,Rigamonti-2015,Byun-2017}. Nevertheless, the account of the long-range tail via a static approximation~\cite{Reining-2002,Botti-2004} \textcolor{black}{typically yields a} single bound exciton peak while being unable to even qualitatively describe multiple excitonic features exhibited by 2D materials, such as monolayers of transition metal dichalcogenides (TMDC), whose optical properties are dominated by several well pronounced bound and continuum excitons~\cite{Mak-2010,Qiu-2013,Ye-2014,Chernikov-2014}. Indeed, \textcolor{black}{it is well recognized that non-adiabatic xc effects are responsible for a number of important physical phenomena exhibited by both finite and extended systems, fostering many attempts to understand the nature of non-adiabaticity in TDDFT and} to construct consistent frequency-dependent approximations~\cite{GroKohn1985,VigKohn1996,DobBunGro1997,VigUllCon1997,TokPan2001PRL,TokStuPan2002PRB,Sole2003,Maitra2004,Barth2005,Fuks2018,Romaniello2009,Botti-2005,UllTok2006PRB,Hellgren2009,Thiele2014,Entwistle2019,Woods2021}.

The central object of the linear response TDDFT is the dynamic xc kernel $f_\textrm{xc}(\omega,{\bf r},{\bf r}'),$ which is responsible for
all interaction effects beyond the random phase approximation (RPA). The exact xc kernel is formally equal to the difference between inverses of the KS density response
function $\chi_{s}(\omega\text{,}{\bf r},{\bf r}')$ and the exact
\textcolor{black}{irreducible} density response function \footnote{\textcolor{black}{As usual, the irreducible response function describes the response to the total field, that is, it excludes the direct Coulomb/RPA contribution.}} $\tilde{\chi}(\omega\text{,}{\bf r},{\bf r}')$,
that is $f_\textrm{xc}(\omega,\rr,\rr')=\chi_{s}^{-1}(\omega,\rr,\rr')-\tilde{\chi}^{-1}(\omega,\rr,\rr')$.
Remarkably, in the early years of TDDFT the very existence of the
xc kernel \textcolor{black}{for frequencies above the absorption threshold} was put in doubt. In 1987 Mearns and
Kohn (MK) demonstrated \cite{MeaKohn1987} that for finite noninteracting
systems the density response function at some special frequencies
may have zero eigenvalues, which indicates that there exist time-periodic
external potentials causing no density response. Apparently this implies
the non-invertibility of~$\chi_{s}$ and, therefore, non-existence
of~$f_\textrm{xc}$~\footnote{It is worth noting that the first observation of zero eigenvalues for the interacting response function has been reported only in the last year~\cite{Woods2021}}. Later, it has been recognized that the MK zeroes do not cause problems for TDDFT. As these zeroes are located strictly at
the real axis, the response function is always invertible for physical
causal dynamics driven by potentials switched on at some initial time~\cite{vanLeeuwen2001}. Despite the absence of conceptual difficulties,
the existence of MK zeroes and the corresponding singularities of~$f_\textrm{xc}$ are considered as disturbing features of the formalism~\cite{vanLeeuwen2001,Hellgren2009,Verdozzi2011,TDDFTbyUllrich}, while their practical
importance remains almost unstudied till now~\cite{Woods2021}.

It is usually assumed that zeros of the density response can \textcolor{black}{appear in some exotic situations and only in the case of finite systems, while} they are generically absent in extended systems in the thermodynamic limit (see e.g. the corresponding discussion in Ref.~\cite{TDDFTbyUllrich}). In this paper we show that the MK zeroes are in fact very common in the long wavelength density response
of solids as they are responsible for non-adiabatic excitonic effects in the optical absorption of semiconductors and insulators. Zeroes of the interacting response function always appear at isolated frequencies between energies of optically active excitons. The corresponding poles of~$f_\textrm{xc}$ represent the key nonadiabatic features required for the TDDFT description of multiple excitonic peaks in the optical spectra. 
Here we illustrate a general relation between the MK zeroes, analytic properties of the xc kernel and excitonic effects in solids. 

The paper is organized as follows. In Section II we demonstrate appearance of zeros of the density response in a simple mechanical toy model. In Section III we highlight the important connection between zeros of the response and the xc kernels in TDDFT, suggesting a path to constructing efficient and practically relevant approximations of $f_\textrm{xc}$. In Section IV we illustrate our approach in practice by applying it to optical absorption in 2D Dirac model and constructing fully analytic $f_\textrm{xc}$ capable of recovering the full Rydberg series of excitonic peaks within the TDDFT. In Sections V and VI we demonstrate application of our approach to designing minimalistic frequency-dependent xc kernels sufficient for reproducing the dominant features of optical spectra of the paradigmatic monolayer TMDCs and bulk solids. In Section VII we discuss the possible impact of our work and its future prospects for highly efficient real-time calculations using TDDFT.

%------------------------------------------
\section{Zeros of the response function}
%of a mechanical model}
%------------------------------------------

Let us first illustrate the significance and the formal origin of zeroes in the response function in a simple toy model.  Consider a mechanical system with three nondegenerate eigenmodes $\omega_A<\omega_B<\omega_C$ illustrated by independent mechanical oscillators $A$, $B$ and $C$ in Fig.~\ref{fig-spring} and coupled to a periodic external field~$E(t)=E(\omega)e^{-i\omega t}$ in
and presence of a small damping ${\eta\ll \omega_n}$:
\begin{equation}
\begin{split}
& \ddot{x}_n = - \omega_n^2 x_n -2\eta \dot{x}_n + Z_n E(t), \\
& n=\text{$A$, $B$, and $C$}.
\end{split}
\end{equation}
Here, $Z_n$ play role of charges of the oscillators. In an ideal isolated system the damping $\eta$ can be understood as the adiabatic parameter describing the periodic drive slowly switched on at $t\to -\infty$. Suppose the object of interest is the collective variable, conjugated to the driving field~$E(t)$:
\begin{equation}
P(t)=\sum_n Z_n x_n(t)
\end{equation}
Thus, the quantity $P(t)$ yields the net polarization induced by the external field.
For the linear response function $\chi(\omega)$ defined by
$P(\omega) = \chi(\omega)E(\omega)$ we obtain:
\begin{equation}
\chi(\omega) = -\sum_n \frac{Z_n^2 }{\omega^2 - \omega_n^2 + 2i\omega\eta}.
\end{equation}
The singularities of the response function $\chi(\omega)$ at $\omega_n$ divide the positive real axis (similarly, for the negative frequency axis) into several domains which correspond to qualitatively different dynamical regimes: (i) $\omega<\omega_A$: all modes oscillate in phase with the external drive, (ii) $\omega_A<\omega<\omega_B$: mode $A$ is out of phase, while $B$ and $C$ are in phase with the external drive, (iii) $\omega_B<\omega<\omega_C$: $A$ and $B$ are out of phase and $C$ is in phase with the external drive and (iv) $\omega_C<\omega$: all modes are out of phase with the external drive, see Fig.~\ref{fig-spring}. In the intermediate regimes (ii) and (iii) there are two special points denoted by $\bar\omega_A$ and $\bar\omega_B$ in Fig.~\ref{fig-spring} where the contributions of modes $A$, $B$ and $C$ to the response counterbalance each other -- these are zeros of $\chi(\omega)$ (which coincide with zeros of the $\Re\chi(\omega)$ in case of the vanishing imaginary part) where the net polarization is practically absent. 
Importantly, for real frequencies $\omega$ and finite $\eta$ the response is never {\em exactly} zero while the inverse $\chi^{-1}(\omega)$ is well defined everywhere, which is of course a manifestation of the general statement by van~Leeuwen~\cite{vanLeeuwen2001}. However, by slowing down the switching process, or by waiting sufficiently long after a sudden switch on, one can make the response at the above two special frequencies arbitrary weak. These two special points are the MK zeroes for our toy model.

The formal reason for existence of a zero response is that the number of microscopic degrees of freedom, defining the number of physical resonances, is larger than the dimension of space hosting the collective variable -- in our toy model example, 3 and 1, respectively. It is worth noting that a similar counting argument was used to prove the possibility of zero eigenvalues for the one-particle Green's function in interacting systems \cite{Gurarie2011}. In the specific case of the density response, one can rephrase this differently: the set of transition densities is always overcomplete in the functional space hosting the density variations. Loosely speaking there are more excitations (resonances) than eigenfunctions spanning the space of densities. As a result, several resonances may in general contribute to one eigenvalue, and the out-of-phase dynamics of densities for different resonances will produce zero response at isolated frequencies in exactly the same way as shown in Fig.~\ref{fig-spring}. 

\textcolor{black}{It is worth emphasizing that the presence of resonances in the dynamic response/correlation function does not imply automatically the existence of zero eigenvalues between resonant frequencies. The simplest examples are the density response function in a one-particle system (e.g. the hydrogen atom), or the one-particle Green's function in noninteracting many-particle system. In both cases, in spite of the resonant structure, there are no zero eigenvalues because the "dimension" of the excitation's space that determines the "number" of resonances coincides with the "dimension" of the Hilbert space where the correlation function acts as a linear operator. As a result there is strictly one resonance per eigenvalue, and thus no zeroes. Only by adding more particles in the case of the density response, or switching on the interaction in the case of the Green's function we make the number of excitations larger than the number of the eigenvalues of the correlation function, which opens a possibility for the appearance of zeros. Our toy model is aimed at demonstrating exactly this point, which is apparently very general for any dynamical theory of reduced/collective variables, such as TDDFT.}

Despite its apparent simplicity, the response of this mechanical toy model and the very appearance of this kind of ``invisibility points'' at frequencies where no net response is observed, shed a light on the analytic structure of the dynamical xc kernel which we will discuss in the next section.

\iffigures
\begin{figure}[h!]
\includegraphics[width=3.3in]{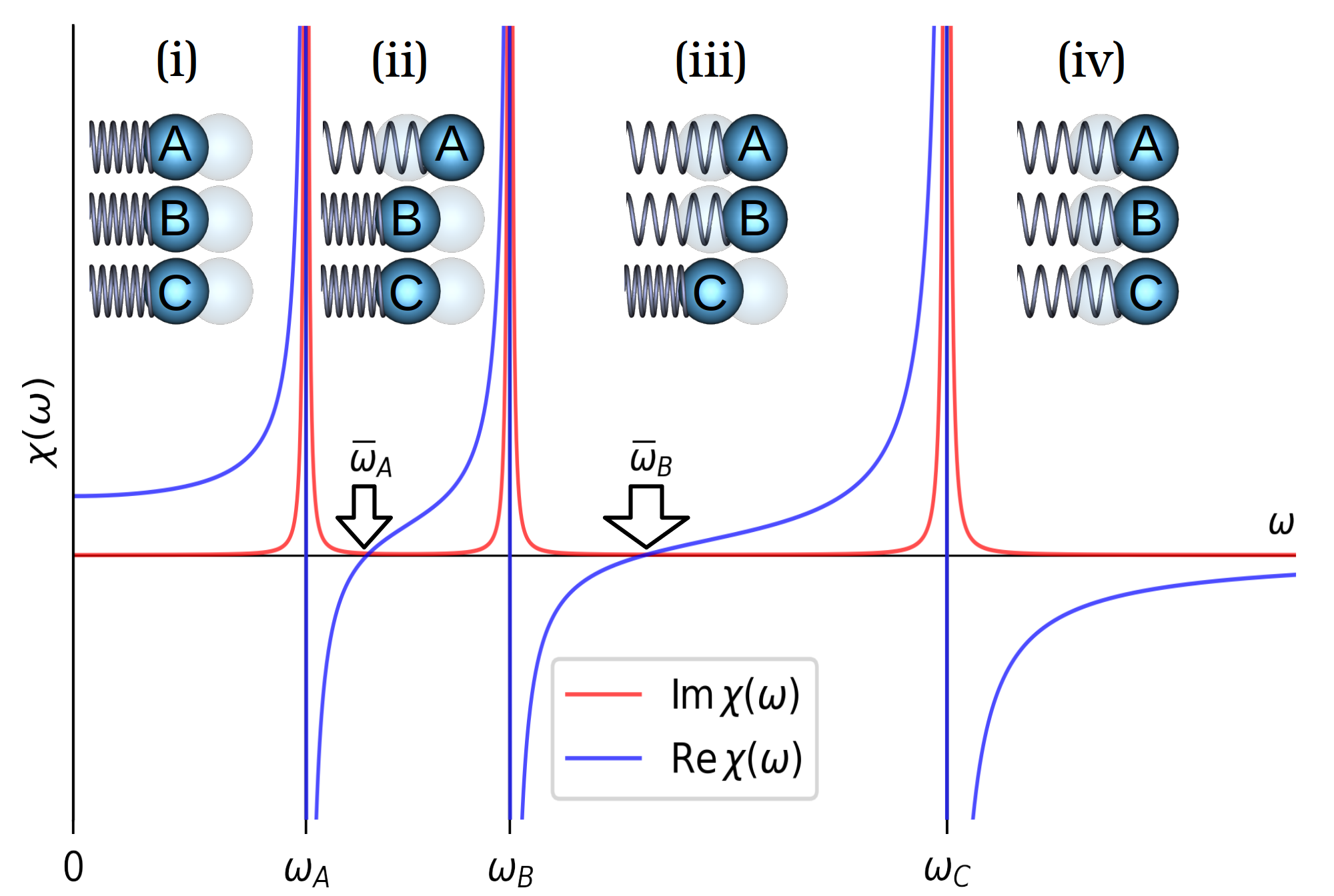}
\caption{\label{fig-spring} 
The real and imaginary parts of the response function $\chi(\omega)$ of a mechanical system possessing eigenmodes $A$, $B$ and $C$ illustrated in the plot as independent oscillators. Arrows mark zeros $\bar\omega_n$ of the response of a collective variable $P(t)$ which arise as a result of the compensated out-of-phase motion of the oscillators. The insets show their relative phases in different dynamical regimes (i)-(iv).
}		
\end{figure}
\fi

%------------------------------------------
\section{Analytic structure of the dynamical long-range xc kernel}
%------------------------------------------
Having discussed the origin of zeros in the response in a simple mechanical model we now focus our attention to many-electron systems. Throughout the paper we will be using Hartree atomic units ($\hbar=m_0=e=1$), unless specified otherwise where expressing the frequencies in units of eV is more natural.
%\begin{equation}
%f_\textrm{xc}(q,\omega) =\chi_S^{-1}(q,\omega)- \tilde\chi^{-1}(q,\omega).
%\label{eq:fxc-def}
%\end{equation}
%where $\chi_S(q,\omega)$ are the Kohn-Sham density-density response function and $\tilde\chi(q,\omega)$ is the proper response
%$$\tilde\chi^{-1}(q,\omega)=\chi^{-1}(q,\omega)+V_C(q)$$
%where $V_C(q)$ is the Coulomb potential.  

Consider a material with the energy gap, such as semiconductor or insulator. \textcolor{black}{In general, in solids the response function $\chi_{\bf{G},\bf{G}'}(\omega,q)$ becomes a matrix in the reciprocal lattice vectors $\bf{G}$. The same is obviously true for the xc kernel. However the excitonic absorption is typically dominated by the head of this matrix (the element with $\mathbf{G}=\mathbf{G}'=0$). Indeed, the head of the density response functions of a gapped material at small momentum $q\to 0$ behaves as $\sim q^2$~\cite{Adler-1962,Wiser-1963}, which implies the famous $1/q^2$ singularity in the head of the xc kernel.} Thus \textcolor{black}{by restricting our attention to the heads of the above matrices}, we can separate the spatial- and frequency-dependent  parts by introducing $\chi(q,\omega) = q^2 \beta(\omega)$, $\chi_s(q,\omega) = q^2 \beta_s(\omega)$ and
$f_\textrm{xc}(q,\omega) = \alpha(\omega)/q^2$ related by
\begin{equation}
\alpha(\omega) =\frac{1}{\beta_s(\omega)}- \frac{1}{\beta(\omega)}.
\label{eq:alpha-def}
\end{equation}
This allows us to focus our attention to the frequency-dependent parts exclusively. In the random phase approximation (RPA) the response function has no singularities 
below the onset of e-h continuum. The account of the exchange and correlation effects results in redistribution of the oscillator strengths and appearance of discrete exciton states with energies $\omega_n$ and oscillator strengths $X_n>0$
inside the fundamental gap $\Delta$:
\begin{equation}
\beta(\omega) = 
\sum_{n} \frac{X_n}{\omega^2-\omega_n^2}
+ \beta^\textrm{reg}(\omega),
\label{eq:chi-additive}
\end{equation}
where the first and second contribution arise from the discrete ($\omega_n<\Delta$) and continuum spectrum, respectively:
\textcolor{black}{we introduced the subscript $\rm reg$ to highlight that the continuum spectrum contribution $\beta^\textrm{reg}(\omega)$ is a regular function with no singularities inside the fundamental gap}.
Because $|\beta^\textrm{reg}(\omega)|<\infty$ for $\omega<\Delta$ and the oscillator strengths, being proportional to the square of the associate transition dipole elements are positive ($X_n>0$), the series of singularities $\{\omega_n\}$ is alternating with series of zeros $\{\bar\omega_n\}$, in direct analogy with the mechanical toy model (cf. Fig.~\ref{fig-spring}).

It has been shown by one of us in Ref.~\cite{Stubner-2004} that the quasiparticle and excitonic parts of the xc kernel can be treated separately. Thus, we assume that the scissor correction has been applied to $\beta_s(\omega)$, so that $\beta_s(\omega)$ and $\beta(\omega)$ have the same gap energy $\Delta$, while focusing on the excitonic contribution only. Because $\beta(\omega)$ enters~\eqref{eq:alpha-def} as the inverse, there arises a correspondence of zeros in the response to singularities of $\alpha(\omega)$. Let us separate these from the xc kernel explicitly. For the imaginary part of $\alpha(\omega)$ we have:
\begin{multline}
\Im\,\alpha(\omega) = 
\Theta(\omega^2-\Delta^2) \,\Im \left[ \frac{1}{\beta_s(\omega)}
- \frac{1}{\beta(\omega)} \right]
\\
+ \sum_{n} \frac{\pi}{\beta'(\bar\omega_n)}\left[
\delta(\omega-\bar\omega_n)-\delta(\omega+\bar\omega_n)
\right],
\label{eq:imfxc}
\end{multline}
where by $\bar\omega_n$ we label the smallest zero higher than $\omega_n$, so that $\omega_n<\bar\omega_n<\omega_{n+1}$. The full complex function $\alpha(\omega)$ is given by a sum of a regular part which has no singularities inside the fundamental gap and a number of discrete poles:
\begin{equation}
\alpha(\omega) =
\alpha^\textrm{reg}(\omega)
+\sum_{n}
\frac{F_n}{\omega^2-\bar\omega_n^2},
\label{eq:fxc}
\end{equation}
with positive oscillator strengths
\begin{equation}
F_n=-\,\frac{2\bar\omega_n}{\beta'(\bar\omega_n)}>0.
\label{eq:Fn}
\end{equation}
The real and imaginary part of $\alpha^\textrm{reg}(\omega)$ are related by the Kramers-Kronig transform and are then defined by:
\begin{equation}
\begin{split}
& \Im\,\alpha^\textrm{reg}(\omega)
= \Theta(\omega^2-\Delta^2) \,\Im \left[ \frac{1}{\beta_s(\omega)}
- \frac{1}{\beta(\omega)} \right],
\\
& \Re\,\alpha^\textrm{reg}(\omega)
= \alpha(\infty) + \frac{1}{\pi}
\mathcal{P}
\int_{-\infty}^{\infty}
\frac{\omega'\,\Im\,\alpha^\textrm{reg}(\omega) }{\omega'^2-\omega^2}d\omega'.
\end{split}
\end{equation}
As seen from Eqs.~\eqref{eq:fxc}-\eqref{eq:Fn}, zeros of the response define the poles of the xc kernel with the strengths given by the slope of the response function at its zeros.

To summarize, Eqs.~\eqref{eq:imfxc}, ~\eqref{eq:fxc} illustrates that the dynamical part of the xc kernel $\alpha(\omega)$ has poles at special frequencies $\{\bar\omega_n\}$ where no response to external perturbation is observed owing to the counterbalanced contributions of the discrete modes. Each of these special frequencies is located strictly between the subsequent pairs of physical excitations.
In the following section we will demonstrate the dynamical xc kernel and the associated pole structure given by~\eqref{eq:fxc} in application to the 2D massive Dirac model.
 
%------------------------------------------
\section{2D massive Dirac model}
%------------------------------------------

Capturing the excitonic effects is one of the long standing difficulties of TDDFT. While the ALDA fails to reproduce excitonic peaks at all, the static long-range corrected (LRC) kernel~\cite{GGG-1997,Reining-2002,Botti-2004,Botti-2007,Sharma-2011,Rigamonti-2015,Byun-2017} with the $\alpha(\omega)$ approximated by a constant $\alpha(\omega)=-\,\alpha_\textrm{static}$, is only capable of capturing single excitonic peak. Although, the attempt to go beyond the static approximation by including a quadratic frequency-dependence in Ref.~\cite{Botti-2005} had demonstrated some improvement in the numerically calculated dielectric function of semiconductors, it is still too simplistic to be applied to 2D semiconductors where excitonic phenomena are dominant features of the absorption spectrum possessing multiple excitonic peaks. In this section we employ the 2D massive Dirac (2DMD) model to illustrate a simple approximation to the dynamical xc kernel arising from the representation~\eqref{eq:fxc} which is capable of capturing not only an isolated exciton peak but the full Rydberg series of excitonic excitations. We consider spinless single valley 2DMD model with Hamiltonian:
\begin{equation}
H_0 = v (k_x\sigma_x + k_y\sigma_y) + \frac{\Delta}{2}\sigma_z,
\label{H0Dirac}
\end{equation}
with interaction between Dirac fermions described by the Rytova-Keldysh potential~\cite{rytova1967the8248,keldysh,Cudazzo2011}:
\begin{equation}
W_\textrm{RK}(q)=-\,\frac{2\pi}{\varepsilon\,q (1+r_0 q)}.
\label{RK}
\end{equation}
Here, $\varepsilon$ is the dielectric constant and $r_0$ is the screening length~\cite{Berkelbach2013,Cho2018}. 
\textcolor{black}{
The single-particle 2DMD~\eqref{H0Dirac} had been shown to appear as a $k.p$ model in TMDC monolayers as a result of interplay between the inversion symmetry breaking and spin-orbit coupling~\cite{Xiao-PRL-2012}. The account of Coloumb interaction into consideration results in the appearance of Rytova-Keldysh potential~\eqref{RK} arising as a result of 2D screening~\cite{rytova1967the8248,keldysh,Cudazzo2011}.   
}

The analytical expression for $\beta_s(\omega)$ \textcolor{black}{follows from the polarization bubble diagram} (see, e.g. Ref.~\cite{Kotov-PRB-2008}):
\begin{equation}
\beta_s(\omega) = \,\frac{v^2}{8\pi \omega}
\left[\frac{\Delta}{\omega}-\left(1+\frac{\Delta^2}{\omega^2}\right)\arctanh{\frac{\omega}{\Delta}}\right].
\label{eq:chi0-analytic}
\end{equation}
The absorption spectra for the 2DMD model in the subgap region are dominated by the $s$-states. Explicit formulas for exciton binding energies of $s$-states in 2D TMDC monolayers were given in the literature~\cite{Olsen-PRL-2016,Molas-PRL-2019, NT-PRB-2022}. 
Employing the effective mass approximation, we use the following semiempirical formula for $s$-states (Eq. (3) of Ref.~\cite{NT-PRB-2022} upon the replacement of $\mu$ by $5\mu/4$~\footnote{The empirical replacement $\mu\to5\mu/4$ is aimed to fix the issue of the Nguyen-Truong~\cite{NT-PRB-2022} systematically underestimating exciton binding energies. Note that the numerical data and theoretical data compared in Fig.2 of Ref.~\cite{NT-PRB-2022} were obtained for different values of the reduced mass which differ by a factor $5/4$ ($\mu=0.20$ in Stier et al.~\cite{Stier-PRL-2018} and $\mu=0.25$ used by Nguyen-Truong). Therefore, Fig.2 of Ref.~\cite{NT-PRB-2022} should be interpreted as the comparison of the numerical data of Stier et al. with the modified version of the Nguyen-Truong formula given by our Eq.~\eqref{eq:NT-formula}. Indeed, our numerical calculations using the BSE confirm that our modified formula~\eqref{eq:NT-formula} produces better values of exciton binding energies in closer agreement with the numerical results than the original Nguyen-Truong formula.}):
\begin{equation}
E_n= -\,\frac{5}{2\mu r_0^2} \left[ \sqrt{n-\frac12+\sqrt{\frac{2\mu r_0}{\varepsilon}}} - \sqrt{n-\frac12}\right]^4.
\label{eq:NT-formula}
\end{equation}
Because the location of zeros of $\chi(\omega)$ is restricted from both sides by adjacent $s$-excitons, a reasonable estimate for $\bar\omega_n$ is provided by extending Eq.~\eqref{eq:NT-formula} to fractional arguments and evaluating it at an intermediate value ${n+\nu}$ between $n$ and $n+1$ with $0<\nu<1$. Thus, we take
\begin{equation}
\bar\omega_n=\Delta + E_{n+\nu}
\label{eq:alpha_n}
\end{equation}
and evaluate it at a central value $\nu=0.5$~\footnote{The numerical analysis using the BSE can improve this result for $\nu$ suggesting a better value of $\nu\approx 0.62$, however, this difference turns out to have a minor impact on the final results of our calculation.}.
Our numerical results suggest that the oscillator strengths $F_n$ follow an \textcolor{black}{empirical} trend
\begin{equation}
F_n \approx 8\varepsilon v \bar\omega_n \Delta
 \left(1-\frac{\bar\omega_n}{\Delta}\right)^{3/2},
\label{eq:dchi}
\end{equation}
which is nearly independent of $r_0$.
Approximating the regular part $\alpha^\textrm{reg}(\omega)$ by a constant
\begin{equation}
\alpha^\textrm{reg}(\omega) \approx
\alpha(\infty)= const,
\label{eq:fxcreg}
\end{equation}
and using Eqs.~\eqref{eq:alpha_n}-\eqref{eq:fxcreg} one obtains a simplest parametrization of $\alpha(\omega)$ for the 2DMD model capable of capturing the full exciton Rydberg series. 

To verify this parametrization we compare the TDDFT response function to the response function obtained by solving the Bethe-Salpeter equation~\footnote{Our code for solving Bethe-Salpeter equation for 2D massive Dirac model is publicly available at~\url{https://github.com/drgulevich/mdbse}.}. 
In the calculations we employed the Tamm-Dancoff approximation which neglects antiresonant terms of the spectra (see, e.g. similar studies in Refs.~\cite{GF-PRB-2011,Wu-PRB-2015}). 
For illustration we take $\varepsilon=1$, $\Delta=2.5\rm\;eV$, $r_0=35\rm\;\angstrom$, $\mu=0.20\,m_0$ which, apart from the spin and valley degrees of freedoms ignored here, are typical for the TMDC monolayers (see e.g. the Supplementary of Ref.~\cite{NT-PRB-2022}). The results for $\Im\chi(\omega+i\eta)$ with broadening $\eta/\Delta=10^{-4}$ are shown in Fig.~\ref{chi-dirac}. The remaining parameter $\alpha(\infty)=-9.7$ was chosen to match the lowest exciton peak to that of the BSE calculation.

\iffigures
\begin{figure}[h!]
\includegraphics[width=3.4in]{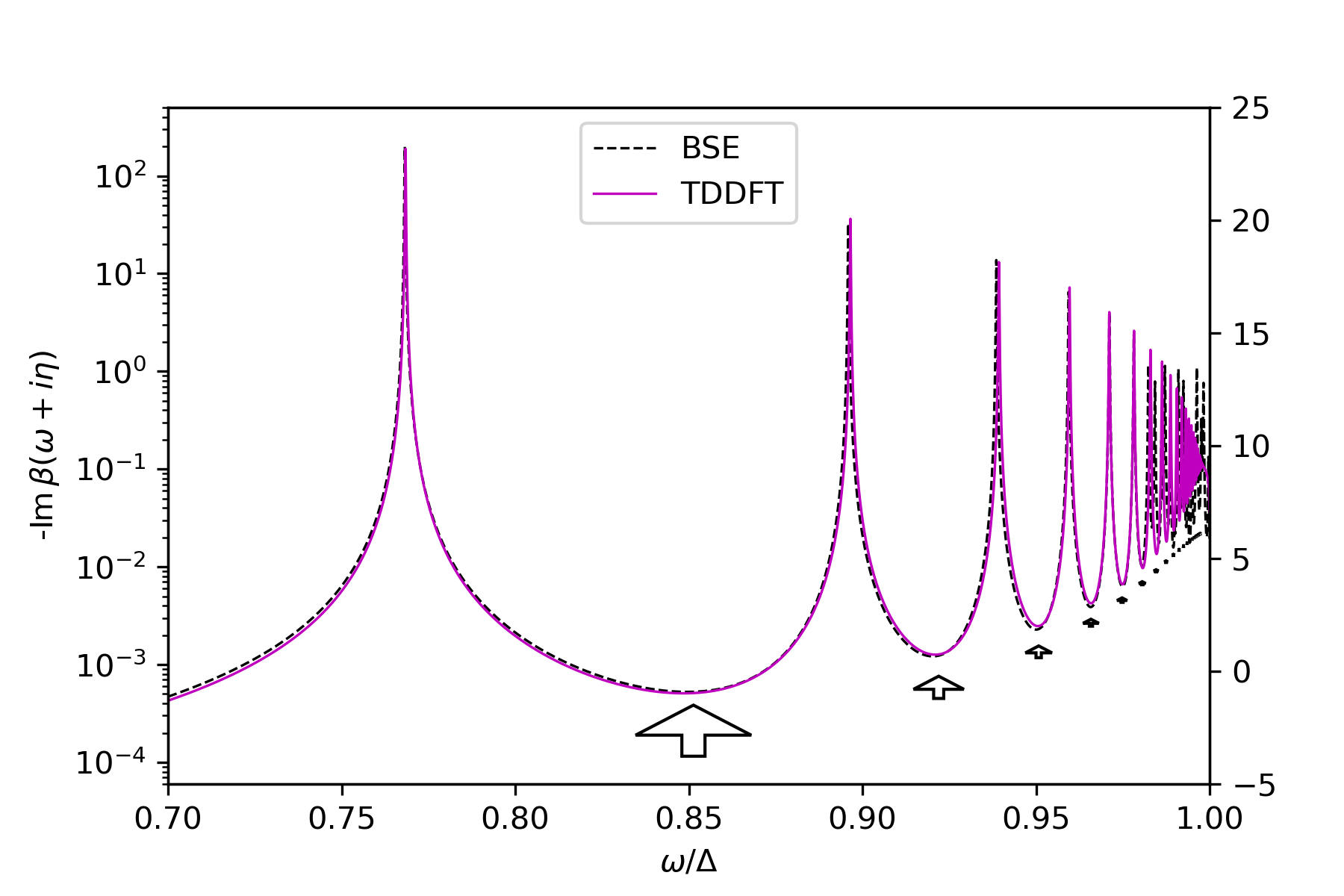}
\caption{\label{chi-dirac} 
Imaginary part of the response function $\beta(\omega+i\eta)$ for the massive Dirac model obtained using our parametrization of the dynamical xc kernel (TDDFT, solid line). For comparison the result of solution of the Bethe-Salpeter equation in Tamm-Dancoff approximation is shown (BSE, dashed line). An artificial broadening $\eta/\Delta=10^{-4}$ was used to smear the singularities. 
Arrows indicate poles of the xc kernel located at zeros of the response $\bar\omega_n$ in accordance with Eq.~\eqref{eq:fxc}. The arrow sizes are scaled with the pole strengths $F_n$. The BSE spectrum is unconverged 
%for $\omega/\Delta \gtrsim 0.98$ 
\textcolor{black}{in the vicinity of the gap due to the finite mesh size and manifests itself as unphysical erratic oscillations at $\omega/\Delta \gtrsim 0.98$.}
%at high frequencies near the gap.
}		
\end{figure}
\fi

%------------------------------------------
\section{2D materials}
%------------------------------------------

In case of ab initio calculations of optical properties of real materials, the high resolution we used to reproduce the Rydberg series in the previous section is redundant. In fact, the experimental absorption spectra of 2D monolayers exhibit few dominant features -- $A$ and $B$ exciton peaks related to the spin-orbit splitting of the Dirac-like dispersion at the $K$ point of the Brillouin zone and a prominent $C$ peak above the quasiparticle gap attributed to excitonic transitions from multiple points around the $\Gamma$ point~\cite{Qiu-PRL-2013,Zhao-ACS-2013}. The account of only these three dominant features returns us to the analogy with the mechanical model of three oscillators discussed in the Section II. In the direct analogy with Fig.~\ref{fig-spring}, there arise two zeros of the response $\bar\omega_A$ and $\bar\omega_B$ which result from the mutual compensation of $A$, $B$ and $C$ excitations and which are of practical importance for reconstructing the xc kernel. Approximating the regular part by a constant $\alpha^\textrm{reg}(\omega)\approx \alpha(\infty)$, Eq.~\eqref{eq:fxc} reads:
\begin{equation}
\alpha(\omega) = \alpha(\infty) +\sum_{n}
\frac{F_n}{\omega^2-\bar\omega_n^2}.
\label{fxc-abc}
\end{equation}
It is common to characterize the excitonic contributions to dielectric function of 2D materials obtained in experiments in terms of Lorentz and Tauc-Lorentz oscillators~\cite{Chernikov-PRL-2014,Funke2016Jul,Diware2017,Ermolaev-JVST-2019,Ermolaev-2020}.
It is therefore practically useful to parametrize the $\alpha(\omega)$ in terms of these quantities.
Recently~\cite{Ermolaev-JVST-2019,Ermolaev-2020} the Tauc-Lorentz parameters for A, B and C peak of MoS$_2$ monolayer were provided. Assuming the frequencies $\omega_n$ and oscillator strengths $X_n$ of $A, B$ and $C$ peaks are given, we can find the poles and their strengths of $\alpha(\omega)$ directly. Because the $C$ peak dominates over $A$ and $B$, the counterbalance points (a.k.a. zeros of $\chi(\omega)$) occur right in the vicinity of the $A$ and $B$ resonances. In this case, zeros can be easily found perturbatively. Denoting by $\beta_n^\textrm{reg}(\omega)$ the regular part of $\beta(\omega)$ with the contribution of the $n$th pole subtracted, the zero $\bar\omega_n$ is defined by:
\begin{equation}
\frac{X_n}{\bar\omega_n^2-\omega_n^2} + \beta_n^\textrm{reg}(\bar\omega_n)=0.
\label{eq:alpha}
\end{equation}
Taking $\beta_n^\textrm{reg}(\bar\omega_n)\approx \beta_n^\textrm{reg}(\omega_n)$ as the zeroth order, $\bar\omega_n$ can then be found iteratively. To find the oscillator strengths we neglect the slowly varying regular part $\beta_n^\textrm{reg}(\omega)$ and obtain:
\begin{equation}
F_n = -\,\frac{2\bar\omega_n}{\beta'(\bar\omega_n)} \approx \,\frac{(\bar\omega_n^2-\omega_n^2)^2}{X_n}
= \frac{X_n}{\beta_n^\textrm{reg}(\bar\omega_n)^2}.
\label{eq:C}
\end{equation}
We use the experimental data from Ref.~\cite{Ermolaev-JVST-2019} for WS$_2$ monolayer on SiO$_2$ substrate and~\cite{Ermolaev-2020} for MoS$_2$ monolayer on SiO$_2$ substrate with perylene-3,4,9,10-tetracarboxylic acid tetrapotassium salt molecule (PTAS), parametrized in the form of Tauc-Lorentz oscillators. From the parameters of oscillators we extract the positions and oscillator strength of the Lorentzian exciton peaks and apply Eqs.~\eqref{eq:alpha} and~\eqref{eq:C} to find the poles and strengths of the dynamical xc kernel summarized in the Table~\ref{table:1}. 

To test the obtained parametrization we used the ab initio code \texttt{exciting}~\cite{exciting-2014}. The band structures of the monolayers were calculated using the LDA functional with account of the spin-orbit coupling. For the linear response calculations we applied the scissor correction 0.54 and 0.82 eV for MoS$_2$ and WS$_2$, respectively, to adjust the fundamental band gap to the experimentally measured values 2.16 eV for MoS$_2$ and 2.38 eV for WS$_2$ monolayers on quartz substrate~\cite{Hill-2016}. As we are interested in the in-plane component of the response, local fields effects were ignored in this study due to their minor effect on the in-plane polarization. The results of our calculation using the TDDFT employing the parametrization from the Table~\ref{table:1} are shown in Fig.~\ref{chi_tddft}.  In the RPA calculations we used artificial broadening $\eta_\textrm{RPA}=0.1\,\textrm{eV}$. In the context of TDDFT, while $\eta_\textrm{RPA}$ is responsible for broadening of the subgap absorption in the continuum, the widths of bound excitons can be tuned by shifting the argument of xc kernel $\alpha(\omega)\to\alpha(\omega+i\eta_\textrm{xc})$. We used $\eta_\textrm{xc}=0.03$ for WS$_2$ and $\eta_\textrm{xc}=0.015$ eV for MoS$_2$. Finally, value of the only remaining parameter $\alpha(\infty)$ affects weakly positions of exciton peaks and was chosen so that to match the resulting peak strengths to the experimentally obverserved values.

\iffigures
\begin{figure}[h!]
\includegraphics[width=3.4in]{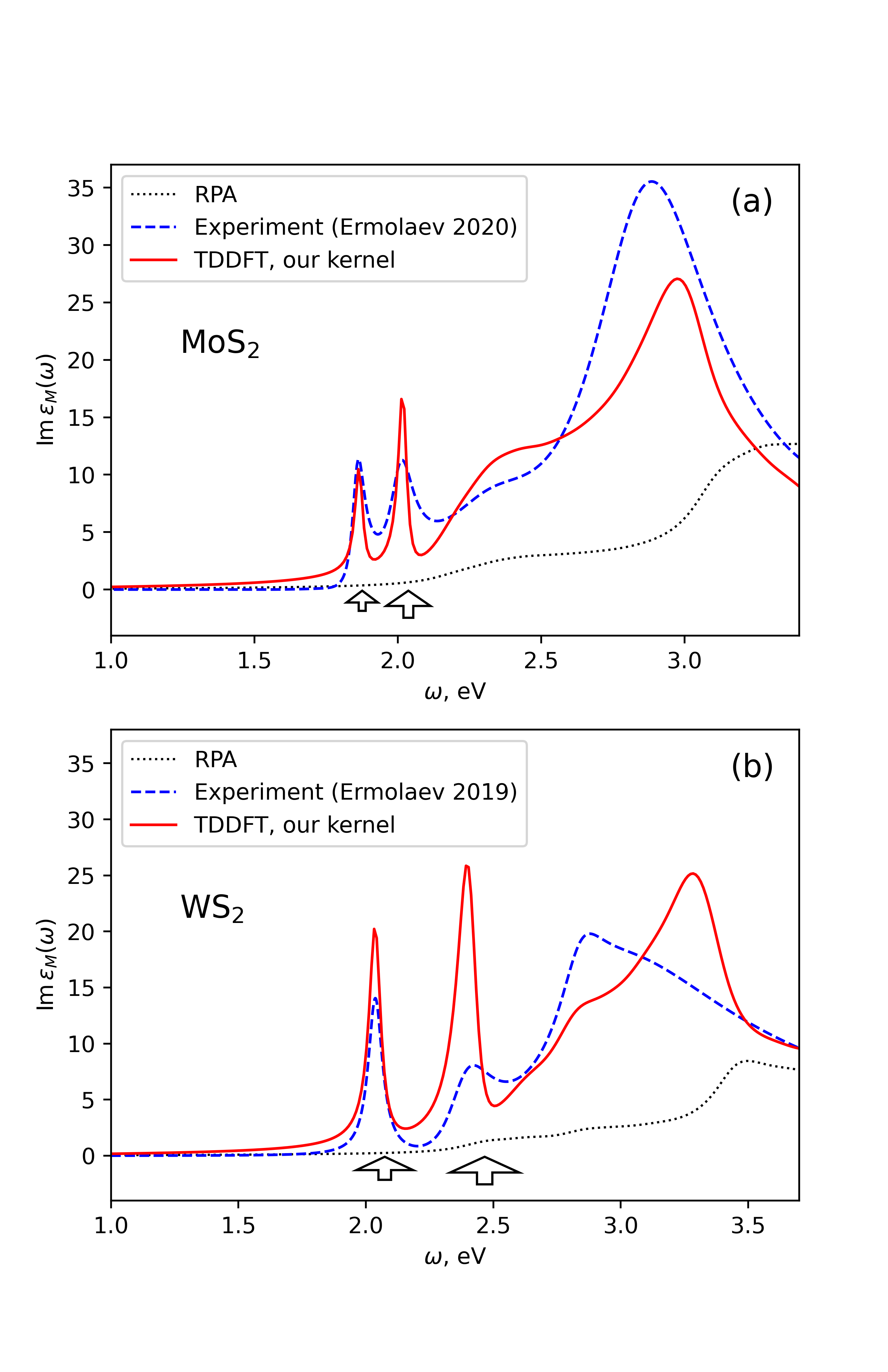}
\caption{\label{chi_tddft} 
Dielectric function of MoS$_2$ (a) and WS$_2$ (b) monolayers calculated using the linear-response TDDFT (black solid curve) with the ab initio \texttt{exciting}
code~\cite{exciting-2014} and the parametrization of the dynamic xc kernel given by the Eq.~\eqref{fxc-abc} with parameters specified in Table~\ref{table:1}. The dashed blue curve marks the experimental data for MoS2 and WS2 monolayers on SiO2 substrate from Refs.~\cite{Ermolaev-2020} and ~\cite{Ermolaev-JVST-2019}), respectively, used to derive the parameters of parametrization. The dotted curve shows the RPA result. Arrows indicate poles of the xc kernel located at zeros of the response $\bar\omega_n$. The arrow sizes are scaled with the pole strengths $F_n$.
}		
\end{figure}
\fi

\begin{table*}[t]
\centering
\begin{tabular}{| c | c | c | c | c | c | c | c | c | c | c | c |} 
 \hline
 Material & $\alpha(\infty)$ & $\bar\omega_1$ & $F_1$ & $\bar\omega_2$ & $F_2$ & $\bar\omega_3$ & $F_3$ & $\bar\omega_4$ & $F_4$ & $\bar\omega_5$  & $F_5$ \\ [0.5ex] 
 \hline
 1L WS$_2$ on SiO$_2$ & $-1.2$ & $2.074$ & $0.125$ & $2.466$ & $0.185$  &  &  &  &  &  & \\  
 1L MoS$_2$ on SiO$_2$  & $-0.9$ & $1.876$ & $0.030$ & $2.037$ & $0.060$  &  &  &  &  &  &  \\
 LiF & $-7.0$ 
 %($-6.6$) 
 & 13.6 & 40.0 &  &  &  &  &  &  &  &  \\  
 Solid Ar  & $-6.7$  & 12.15 & 1.6 & 12.70 & 57.0 & 13.60 & 4.0 & 13.79 & 2.0 & 13.94 & 1.0 \\
 \hline
\end{tabular}
\caption{Parametrization of the dynamical xc kernel for 2D monolayers and bulk materials. 
%The value in brackets for $\alpha(\infty)$ is provided for the case when the local field effects are ignored. 
Values of $\alpha(\infty)$ are given in Hartree atomic units. For convenience, $\bar\omega_n$ and $F_n$ are expressed in eV and eV$^2$, respectively, so that to restore the result for $\alpha(\omega)$ in atomic units.
}
\label{table:1}
\end{table*}

%----------------------------------------------
\section{Bulk solids}
%----------------------------------------------

Similar to the case of 2D materials discussed above, the same approach applies to bulk wide-gap semiconductors and insulators exhibiting bound excitons in the absorption spectra. Some of the illustrative examples are LiF and solid Ar which received a great deal of attention in a range of  theoretical studies~\cite{Rohlfing1998Sep,Sottile2003Nov,Sottile2007Oct,Marini-2003,Galamic-2005,Ullrich2016,Byun-2017}. 

Experimental absorption spectrum of LiF exhibits a pronounced excitonic peak at 12.6 eV followed by an unresolved Rydberg series~\cite{Piacentini1975Sep} and a featureless quasiparticle band gap at 14.1-14.2 eV~\cite{Shirley1996Apr}, see Fig.~\ref{lif-ar}a. We therefore should expect the zero of the response function to appear between the first exciton and the onset of the Rydberg series at 13.6 eV. By placing a pole at $\bar\omega_1=13.6$ eV we obtain a simple one-pole parametrization of the xc kernel for LiF presented in the Table~\ref{table:1}. The results obtained using the static~\cite{Reining-2002,Botti-2004}
and parabolic~\cite{Botti-2005} LRC kernels are also shown in Fig.~\ref{lif-ar}a for comparison~\footnote{We have modified values of the parameters used in Ref.~\cite{Botti-2005} to align the central exciton peak frequency with that of the BSE result. In Fig.~\ref{lif-ar}a we used $\alpha(\infty)=-8.4$ for the static and $\alpha(\omega)=-1.5-0.042(\omega/{\rm eV})^2$ for the parabolic LRC kernel.}. Presence of the pole redistributes the oscillator strengths towards the higher energies causing a secondary peak at 14.5 eV, which is present in BSE and experiment but absent in the static and parabolic LRC approaches. Moreover, in contrast to the parabolic LRC, our xc kernel does not suppress the absorption at higher energies, as seen in Fig.~\ref{lif-ar}a. For the ease of comparison with BSE calculations we used the same broadening $\eta_\textrm{xc}=\eta_\textrm{RPA}=0.25$~eV as used in Ref.~\cite{Rohlfing1998Sep}. 

The absorption spectrum of solid Ar consists of a series of well separated exciton peaks~\cite{Saile-1976,Saile1980Dec,Bernstorff1986Jun}. In the same fashion, with account of the dominant $n=1$, $1'$, $2$, $2'$ and $3$ excitonic features, we obtain the parametrization for xc kernel presented in Table~\ref{table:1}. The dielectric function calculated results using this parametrization is presented in Fig.~\ref{lif-ar}b.
%(calculations with and without the account of the local fields visually coincide). 
We used arificial broadenings $\eta_\textrm{RPA}=0.25$~eV and $\eta_\textrm{xc}=0.03$~eV in RPA and $\alpha(\omega)$, respectively. It is interesting to note that the two lowest exciton states $n=1$, $1'$ are well separated from the rest of the spectra. \textcolor{black}{In this case of two isolated excitonic peaks the analytic structure of $f_{xc}(\omega)$ has a remarkable analogy to that for double excitations in finite systems, where a pole in the frequency dependence of the xc kernel is shown to appear~\cite{Maitra-2004,Gritsenko-2009,Woods2021}. In both systems the pole in $f_{xc}(\omega)$ reflects the presence of two "non-orthogonal" excitation channels, which interfere destructively, leading to the suppression of the response at some isolated frequency.}

\iffigures
\begin{figure}[h!]
\includegraphics[width=3.4in]{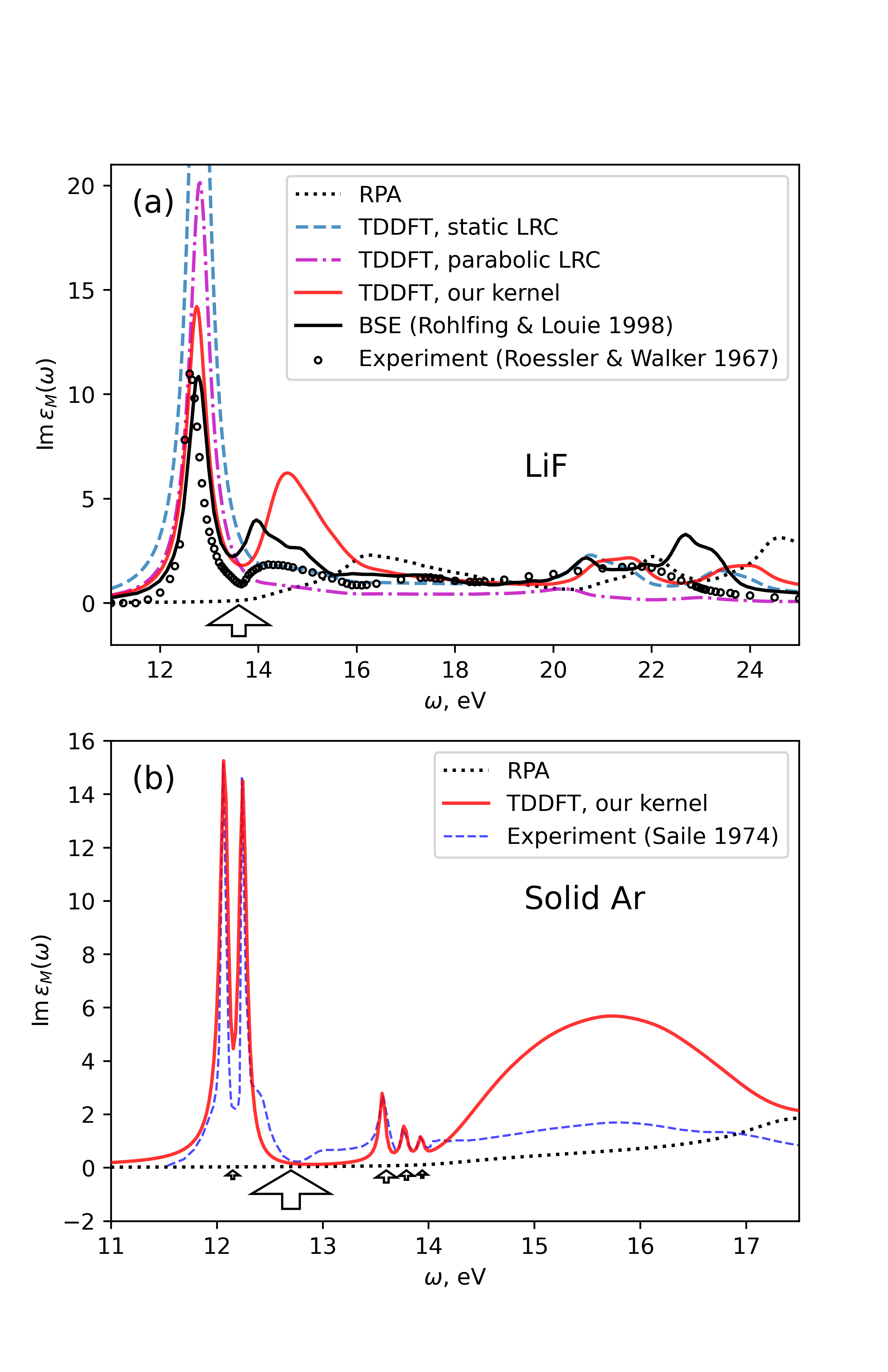}
\caption{\label{lif-ar} 
Imaginary part of the dielectric function for LiF (a) and solid Ar (b). Aritificial homogeneous broadening has been used $\eta_\textrm{xc}=\eta_\textrm{RPA}=0.25$~eV (same for $\alpha(\omega)$ and RPA) for LiF  to match the broadening in the BSE result~\cite{Rohlfing1998Sep}. The broadenings for solid Ar are: $\eta_\textrm{xc}=0.03$~eV and $\eta_\textrm{RPA}=0.25$~eV. The experimental data is taken from Refs.~\cite{LiF-1967} and Ref.~\cite{Saile-PhD} for LiF and Ar, respectively. Arrows indicate locations of the poles of the xc kernel at zeros of the response $\bar\omega_n$. The arrow sizes are scaled with the pole strengths $F_n$.
}		
\end{figure}
\fi

%----------------------------------------------
\section{Discussion}
%----------------------------------------------
\textcolor{black}{We demonstrate that zero eigenvalues of the density response function, discovered many years ago by Mearns and Kohn for model systems, represent a common feature of the optical absorption in insulating solids. In fact, they dominate the frequency dependence of the xc kernel of TDDFT, especially when several excitonic peaks, both bound and continuum, are present in the absorption spectra. The latter is especially relevant for insulating materials of reduced dimension, such as TMDC. Shedding the light on the nature of non-adiabatic effects in the TDDFT of excitonic absorption, which are characterized by the poles of the dynamical xc kernels} allows to design simple and practically efficient approximations of the kernels for the ab initio description of the collective many-body phenomena within the TDDFT. Our approach illustrated for TMDC monolayers and bulk materials hints the procedure for obtaining simple and practical dynamical xc kernels for a variety of semiconducting and insulating materials directly from experimental absorption spectra. The algorithm for reconstructing xc kernels can look as follows:
(i) Experimentally obtained imaginary part of the macroscopic dielectric function is parametrized in terms of Lorentz oscillators.
(ii) With the peak widths of the oscillators discarded, the central frequencies and oscillator strengths used to find poles and oscillator strengths of the xc kernel with the help of Eqs.~\eqref{eq:alpha} and~\eqref{eq:C}.
(iii) The regular part of the dynamical xc kernel is approximated by a constant $\alpha^\textrm{reg}(\omega)\approx \alpha(\infty)$.

Apart from the physical insight, xc kernels in the form~\eqref{fxc-abc} provide a natural tool for the highly efficient practical implementation of the real-time TDDFT. Indeed, while the approach~\cite{Ullrich-PRL-2021} summarizes the scheme for the calculation using the static LRC kernel, the pole expansion used in Eq.~\eqref{fxc-abc} enables to extend this further to deal with the frequency-dependent kernels with the help of a highly efficient numerical approach proposed in Ref.~\cite{Gulevich-PRB-2019}, where the performance comparable to that of the standard time-dependent ALDA has been demonstrated for the case of Vignale-Kohn functional~\cite{Vignale-1996,VUC-PRL-1997}. Based on this, the perspectives of the real-time ab initio dynamics with the help of TDDFT and the account of excitonic effects thus emerge at a full scale, which yet in the case of TMDC monolayer alone is of a high practical and fundamental interest. The real-time calculations with TDDFT with the proper account of excitonic phenomena may shed a new light on the intriguing nonlinear phenomena~\cite{RT-GW-BSE,Yaroslav-PRB-2022} while paving the way to a practical impact in semiconducting physics, materials science and optoelectronics.

\section*{Acknowledgements}
The authors would like to thank Prof. Volker Saile for helpful clarifications of the experimental data for solid~Ar. \textcolor{black}{Numerical calculations with TDDFT by D.R.G. were supported by the Russian Science Foundation under the grant 18-12-00429. Analytical calculations for the Dirac model by V.K.K. were supported from the Georg H. Endress foundation. Theoretical insights and analytical calculations by I.V.T. related to the connection between $f_{xc}$ poles and Mearns-Kohn zeroes were supported by Grupos Consolidados UPV/EHU del Gobierno Vasco (Grant IT1453-22) and by the grant PID2020-112811GB-I00 funded by MCIN/ AEI/10.13039/501100011033}.

%\bibliography{refs}


\end{document}